\documentstyle[12pt,psfig]{article}
\begin{document}

\title{{\bf Limitations of the $\Phi$ measure of fluctuations in
event-by-event analysis}}
\author{O.V.Utyuzh$^{1}$\thanks{%
e-mail: utyuzh@fuw.edu.pl},~ G.Wilk$^{1}$\thanks{%
e-mail: wilk@fuw.edu.pl} and Z.W\l odarczyk$^{2}$ \thanks{%
e-mail: wlod@pu.kielce.pl} \\
$^1${\it The Andrzej So\l tan Institute for Nuclear Studies}\\
{\it Ho\.za 69; 00-689 Warsaw, Poland}\\
$^2${\it Institute of Physics, Pedagogical University}\\
{\it Konopnickiej 15; 25-405 Kielce, Poland}}
\date{\today}
\maketitle

\begin{abstract}
We provide a critical overview of the $\Phi$ measure of fluctuations
and correlations. In particular we show that its discriminating power
is rather limited in situations encountered in experiment.\\

\end{abstract}

\newpage

It is widely recognized that event-by-event analysis of experimental
data on multiparticle reactions (and especially studies of
fluctuation patterns seen there) provides us with very important and
sensitive tool in our attempts to understand dynamics of heavy ion
collisions \cite{GEN}. They are particularly useful in searching for
some special features of the quark-gluon plasma (QGP) equation of
state \cite{QCD}. The question of the best method of their
investigation allowing for the most information to be gathered is
therefore of great interest (cf., for example, reviews \cite {TATHH}
and references therein). Some time ago, a novel method of
investigation of fluctuations in even-by-event analysis of high
energy multiparticle production data was proposed and applied to
nuclear collisions \cite{GM}. It is based on a suitably defined measure
$\Phi$,
\begin{equation}
\Phi_x\, =\, \sqrt{\frac{\left\langle Z^2\right\rangle}{\langle N\rangle}}\,
-\, \sqrt{\bar{z^2}} \qquad {\rm where}\qquad Z\, =\, \sum^N_{i=1}\, z_i ,
\label{eq:FI}
\end{equation}
which originally was supposed to discriminate wheather or not
fluctuations of a given observable $x$ is exactly the same for
nucleon-nucleon and nucleus-nucleus collisions \cite{FOOT}.
Here $z_i = x_i - \bar{x}$ where $\bar{x}$ denotes the mean value of
the observable $x$ calculated for all particles from all events (the
so called inclusive mean) and $N$ is the number of particles analysed
in the event. In (\ref{eq:FI}) $\langle N\rangle$ and $\langle
Z^2\rangle$ are averages of event-by-event observables over all
events, whereas the last term is the square root of the second moment
of the inclusive $z$ distribution. By construction $\Phi_x = 0$ for
independently produced particles \cite{GM}.\\

However, application of this method is not free from controversy.
When first applied to NA49 data for central $Pb-Pb$ collisions at
$158$ A$\cdot$GeV \cite{R} it apparently revealed that fluctuations
of transverse momentum ($x=p_T$) decreased significantly with respect
to elementary NN collisions. This in turn has been interpreted as a
possible sign of equilibration taking place in heavy ion collisions,
providing thus an enviroment for the possible creation of QGP. It was
immediately realised that existing models of multiparticle production
are leading in that matter to conflicting statements \cite{ANALYS}.
The more recent NA49 data \cite{DATA} reported, however, a new value
(almost an order of magnitude greater then the previous one), which
was the one corresponding to a pion gas in global equilibrium \cite{M}.
A number of attempts followed, trying to clarify the meaning of
$\Phi$ (cf., for example, \cite{BK,OTHERS} and references therein.)
In the mean time, it was extended to study event-by-event fluctuations
of "chemical" (particle type) composition of produced secondaries
\cite{G}, to study azimuthal correlations among them (which are
important for studies of flow patterns observed in heavy ion
collisions \cite{SMAPP}) and to cover also higher order correlations
\cite{SMMULT}. Finally, $\Phi$ has been also analysed by means of the
nonextensive statistic both for $p_T$ correlations \cite{FLUQ} and
for fluctuations of chemical composition as well \cite{UWW}.\\

In \cite{UWW} we have pointed that, if there are some additional
fluctuations (not arising from quantum statistics, like those caused
by the experimental errors), which add in the same way to both terms
in definition (\ref{eq:FI}) of $\Phi $, it would perhaps be better to
use another form of it, like for example
\begin{equation}
\Phi\, \rightarrow \, \Phi ^{\star }\, =\,
       \frac{\left\langle Z^{2}\right\rangle }{\langle N\rangle }
       -\bar{z^{2}},  \label{eq:PHISTAR}
\end{equation}
where the variances $\sigma_x^2$ would cancel (being present only
implicitly). We would like to elaborate on this criticism towards
$\Phi$ (as well as $\Phi^{\star}$) measure in more detail here. Our
point is (cf. also \cite{TATHH}) that inclusive experiments provide us
both with single particle distributions $P(x)$ (i.e., with
information on fluctuations of the $x$-values) and with multiplicity
distributions $P(N)$ (i.e., with information on fluctuations of $N$
and, when put together with $P(x)$, also with information on
correlations between $x$-value and multiplicity $N$). Also
correlations between produced particles (i.e., between $x$-values),
especially Bose-Einstein correlations (BEC) resulting from their
statistics, are to a large extend known \cite{BEC}. Measures $\Phi$ (or
$\Phi^{\star}$) depend on all of them, because (notice that
$\bar{z^{2}}=\sigma _{z}^{2}=\sigma _{x}^{2}$) 
\begin{equation}
\langle Z^{2}\rangle \,=\,\sigma _{Z}^{2}\,=\,\langle N\rangle \sigma
_{x}^{2}\,+\,\langle N(N-1)\rangle \cdot
    {\rm cov}(x_{i},x_{j})\,+\,c(x,N) , \label{eq:Z}
\end{equation}
where covariance ${\rm
cov}(x_{i},x_{j})=\overline{x_{i}x_{j}}-\bar{x}_{i}\bar{x}_{j}=
\rho \sigma _{x}^{2}$, i.e., is given in terms of the variance
$\sigma _{x}^{2}$ of the variable $x$ and their mutual correlation
coefficient $\rho $, while the last term describes the possible
correlation between the variable $x$ and the multiplicity $N$. Notice
that assuming, for example, $\langle x\rangle _{N}$ for given $N$
being given by 
\begin{equation}
\langle x\rangle _{N}\,=\,\bar{x}\left[ 1\,+\,\alpha \cdot \frac{N-\langle
N\rangle }{\langle N\rangle }\right]   \label{eq:xN}
\end{equation}
one gets for correlation term 
\begin{equation}
c(x,N)\,=\,\alpha ^{2}\cdot \bar{x}^{2}\cdot \left( \langle N^{2}\rangle
-\langle N\rangle ^{2}\right) .  \label{eq:defc}
\end{equation}
Therefore $\Phi=0$ is achieved only for independent $x$-values, in
which case they are also uncorrelated. If this is not the case 
they are given by:
\begin{eqnarray}
\Phi\, &=&\, \sqrt{\sigma^2_x\, +\, \frac{\langle N(N-1)\rangle}{\langle
N\rangle} \cdot {\rm cov}(x_{i},x_{j})\, +\, \frac{\langle N^2\rangle -
\langle
N\rangle^2} {\langle N\rangle}\alpha^2 \bar{x}^2} \, -\, \sqrt{\sigma_x^2}
\nonumber\\
 &=&\, - \sigma_x\, +\, \sqrt{\Phi^{\star} + \sigma_x^2} , \label{eq:full}\\
\Phi^{\star }\, &=&\,\left( \,\langle N\rangle - 1 + \,
                    \frac{\sigma _{N}^{2}}{\langle N\rangle }\right)
                    \cdot {\rm cov}(x_{i},x_{j})\, + \,
                    \frac{\sigma _{N}^{2}}{\langle N\rangle }
                    \alpha ^{2}\bar{x}^{2}\nonumber\\
                   &=&\,
                    \Phi (\Phi\, +\, 2\sigma_x).  \label{eq:STAR}
\end{eqnarray}

Because in reality (neither in experiments nor in models or event
generators attempted to describe them) conditions of independence are
not fulfilled, $\Phi,~\Phi^{\star} \neq 0$ and a question arises: what
information do they convey? In particular, can we learn more from
event-by-event analysis performing it by means of the $\Phi$ (or
$\Phi^{\star}$) measure rather than by being satisfied with the
known inclusive distributions alone? In this respect one should
notice that \cite{TATHH}: 
\begin{itemize}
\item Both $\Phi $ and $\Phi^{\star}$ depend strongly on the
multiplicity fluctuations and this dependence vanishes only
in the limiting case of a Poisson distribution (for which $\langle
N^{2}\rangle -\langle N\rangle ^{2}=\langle N\rangle $ and 
$\langle N(N-1)\rangle =\langle N\rangle ^{2}$) where, for example,
\begin{equation}
\Phi ^{\star }\,=\,\langle N\rangle \cdot {\rm cov}(x_{i},x_{j})\,+
\,\alpha^{2}\bar{x}^{2}.  \label{eq:Pois}
\end{equation}
\item Experimental data allow us to study the dependence of $\langle
x\rangle _{N}$ on the multiplicity $N$ in an explicit way.
In the case when such correlations are absent, the $\Phi ^{\star }$
measure reduces simply to a two-particle correlation measure
\begin{equation}
\Phi ^{\star }\,=\,\left( \langle N\rangle -1\,+\,
 \frac{\sigma _{N}^{2}}{\langle N\rangle }\right)
 \cdot {\rm cov}(x_{i},x_{j}).  \label{eq:indep}
\end{equation}
\item  The influence of the two-particle correlation term ${\rm
cov}(x_{i},x_{j})$ depends on the kind of particles involved: $\Phi
=0$ for Boltzman statistics, $\Phi <0$ for fermions and $\Phi >0$ for
bosons. However, this type of correlations, especially for BEC for
bosons, are subject to extensive experimental and theoretical
investigations where one mesures or modells the so called
$2-$particle correlation function $C_{2}(x_{i},x_{j})$ \cite{BEC}.
Because
\begin{equation}
{\rm cov}(x_{i},x_{j})\,=\,\int \int
\,dx_{i}dx_{j}\,x_{i}\frac{dn}{dx_{i}}%
\,x_{j}\frac{dn}{dx_{j}}\,\left[ C_{2}(x_{i},x_{j})\,-\,1\right] ,
\label{eq:C2}
\end{equation}
it means that $\Phi$ and $\Phi^{\star}$ measures are
nothing but the correlation measure $C_{2}$ averaged over 
single particle distributions $dn/dx$ (i.e., we are in fact not
gaining but losing some information contained in $C_2$ and $dn/dx$
and  none of $\Phi$'s provide us with any new information).
\end{itemize}

The above remarks, being obvious and essentially known,
\cite{TATHH,BK}, have so far not been backed by any convincing 
numerical illustration. By this we mean calculations using an event
generation algorithm which would satisfy all conservation laws
(especially energy-momentum conservation) and at the same time model
also the BEC. Because both points seem to play a major role in a proper
description of $\Phi$ (or $\Phi^{\star}$) measure \cite{TATHH,DATA},
which was not checked properly \cite{FOOT1}, we would like to fill
this gap by calculating $\Phi$ and $\Phi^{\star}$ in simple models of
hadronization using an algorithm which preserves both energy-momentum
conservation and the original single-particle distributions and
models at the same time all features of BEC in such processes
\cite{TIHANY}. To this end we shall compare $\Phi$ and $\Phi^{\star}$
measures as function of mean multiplicity $\langle N^{(-)}\rangle$ of
negatively charged particles for hadronization processes proceeding
without and with BEC. Two simple models of hadronizations of mass $M$
will be considered: the cascade model (CAS) developed by us recently
\cite{CAS} (where the whole space-time and phase-space history of the
hadronization process is explicitly known) and simple statistical
model (MaxEnt) based on information theory approach proposed in
\cite{MaxEnt} (where details of hadronization are not available; both
models were already used in \cite{TIHANY}).\\

Most of the above mentioned applications of $\Phi$ measure concerned
transverse momenta $x=p_T$ \cite{GM,R,M,DATA}, which is always
positive. The corresponding variable here would be $|p|$
\cite{FOOT2}, results for which are shown in Fig. 1. We have found it
interesting to enlarge analysis for the variable which can take any
sign, as momentum $p$ in our case, results for which are shown in
Fig. 2. The differences are striking. Whereas behaviour of
$\Phi(x=|p|)$ is similar to that for $x=p_T$, that of $\Phi(x=p)$ is
much more sensitive to the limits imposed by the energy-momentum
conservation (both measures are negative here). In both Figs. 1 and 2
we present $\Phi$ and $\Phi^{\star}$ for CAS and MaxEnt models
without and with BEC, in this later case for two different choices of
the weights specifying BEC \cite{FOOT3}: constant $P=0.5$ and
Gaussian $P$ as discussed in \cite{TIHANY} (which lead to very
different BEC patterns). The results are given as functions of the
mean multiplicity $\langle N^{(-)}\rangle$ of negatively charged 
secondaries produced  in hadronization process of mass $M$.\\ 

The special feature emerging from our calculation is the fact that
the effect caused by BEC is clearly visible in all cases (it is maximal
for the case of constant weights because the BEC is maximal there
\cite{TIHANY}). To make dependence on BEC more clear we present in
Fig. 3 for $x=|p|$ case the corresponding differences $\delta \Phi =
\Phi_{BEC} - \Phi_{noBEC}$ and $\delta \Phi^{\star} =
\Phi_{BEC}^{\star} - \Phi_{noBEC}^{\star}$. The curves show the best
power fits of the type $\delta \Phi \propto \langle
N^{(-)}\rangle ^{\delta}$ exhibiting very strong and growing
dependence of the effect of BEC on $\langle N^{(-)}\rangle$. There
are some other characteristic features of the results worth
attention, namely: $(i)$ the scale of effect is different in $\Phi$ 
and $\Phi^{\star}$ this reflects differences in their definitions
(\ref{eq:full}) and (\ref{eq:STAR}); $(ii)$ their different
dependencies on $\langle N^{(-)}\rangle$ indicate that $\sigma_x$
plays a significant role here; $(iii)$ the completely different
behaviours for $x=p$ and $x=|p|$ (compare Fig. 1 with Fig. 2)
illustrate how $\Phi_x$ changes dramatically whether the measured
variable is restricted to being positive or not. Because in
the second case the role of the energy-momentum constraints is much more
important, dependence on this constraint is more pronounced. Notice
that points for MaxEnt in Figs. 1 and 2 are consistently higher than
for CAS. This should probably be attributed to different ways of
imposing conservation laws in both models: they are satisfied
globally in the statistical model MaxEnt and locally, in every branching
point, in the cascade model CAS. Because of this CAS prefers symmetric
distributions in momenta (here forward-backward) whereas MaxEnt
allows more frequently for asymmetric ones (for example, when one
particle with large momentum is balanced by a bunch of low momenta
particles in other hemisphere). The difference between both types of
models seen in Fig. 3 originate from different shapes of
$C_2(p_i,p_j)$ functions, which are broader (what corresponds to
smaller "radius" $R$) for CAS type of models.\\

To conclude: we have demonstrated using numerical algorithm which
preserves energy-momentum conservation and at the same time models
BEC in multiparticle hadronization that $\Phi$ (and also
$\Phi^{\star}$) measure is very sensitive both to the constraints
provided by conservation laws and to the effects of correlations
(exemplified here by the BEC). Both features must therefore be 
carefully accounted for when attempts are made to reach some
conclusions concerning new physical effects when using these measures
to the event-by-event analysis of data. We have also written down
explicit relations between both measures and all other observables
(well defined and known in statistics) available from inclusive
experiments. They show explicitly that both $\Phi$'s are closely
related to inclusive correlation functions. In this respect we believe 
that no new information is obtained from these measures in comparison
with what is available from inclusive measurements.\\

The partial support of Polish Committee for Scientific Research
(grants 2P03B 011 18 and  621/E-78/ SPUB/ CERN/ P-03/ DZ4/99) is
acknowledged.

\newpage
\noindent
{\bf Figure Captions:}

\begin{itemize}

\item[{\bf Fig. 1}] $\Phi$ (in GeV, left panels) and $\Phi^{\star}$ 
                    (in GeV$^2$, right panels) as function of mean 
                    (negatively charged) multiplicity 
                    $\langle N^{(-)}\rangle$ calculated for 
                    $x=|p|$ for two different hadronization models, 
                    CAS and MaxEnt, without and with BEC. 
                    Upper panels contain BEC obtained using constant 
                    weights, lower panels contain BEC corresponding 
                    to Gaussian weights \cite{FOOT3}. 
                    Black symbols denote CAS and MaxEnt
                    models with BEC, open symbols without BEC. 
                    Stars correspond to MaxEnt and circles to CAS
                    models, respectively. Lines are just interpolating
                    between calculated points.

\item[{\bf Fig. 2}] The same as in Fig. 2 but this time for $x=p$.

\item[{\bf Fig. 3}] Differences between events with and without BEC
                    for $x=|p|$, $\delta \Phi$ (upper part) and 
                    $\delta \Phi^{\star}$ (lower part) 
                    for models and BEC weights presented in Fig. 1
                    (with the same meaning of different symbols).
                    Circles denote CAS and stars MaxEnt
                    models with, respectively, constant (for open 
                    symbols) and Gaussian (for full ones) weights
                    for BEC. Lines indicate attempts of best power-like
                    fits of type $\delta \Phi \propto \langle
                    N^{(-)}\rangle ^{\delta}$
                    with $\delta$ equal to (going from top) to
                    $2.0,~2.0,~1.5,~0.7$ for upper panel and
                    $3.6,~3.6,~3.1,~2.2$ for lower panel.

\end{itemize}

\newpage

\begin{figure}[h]
\psfig{file=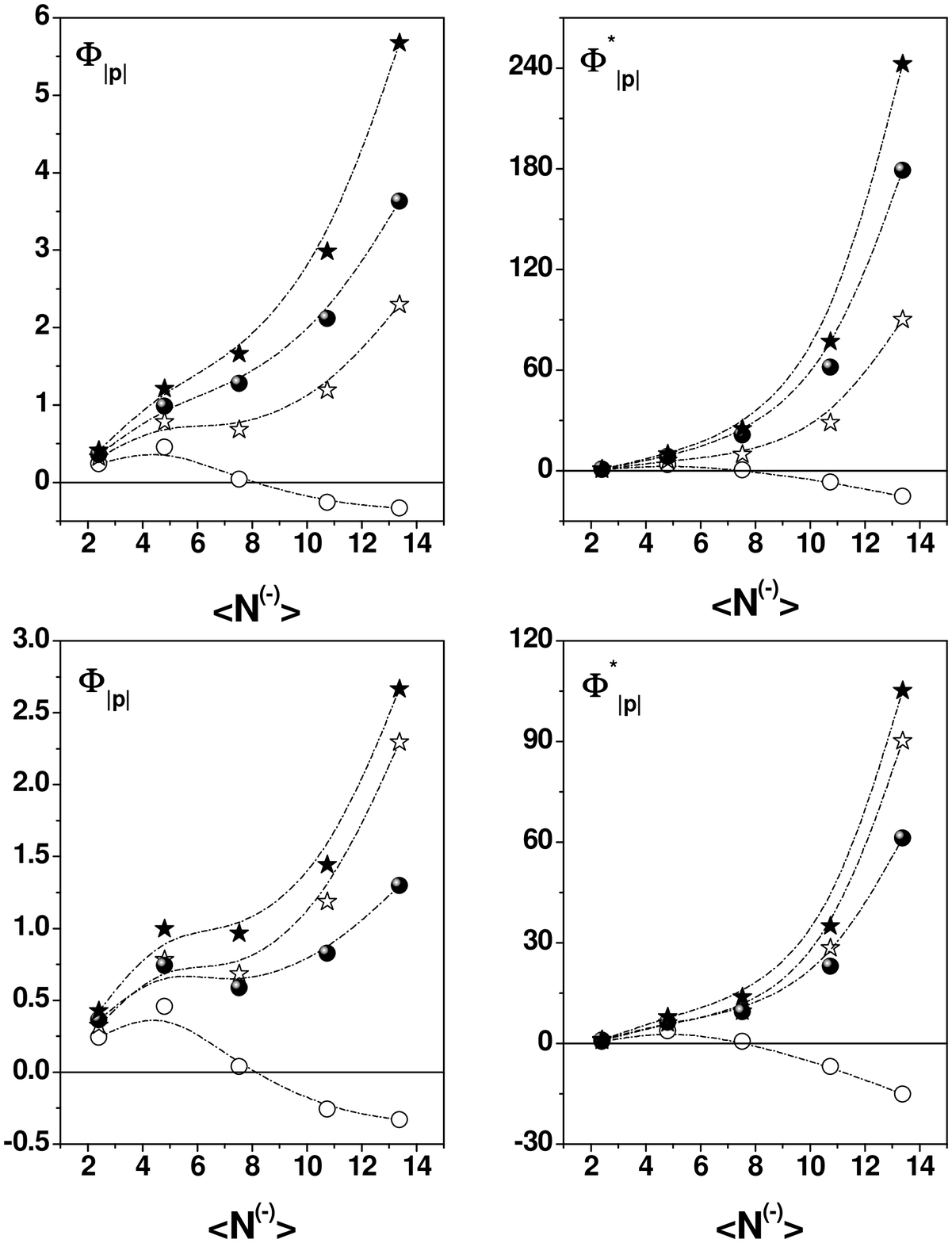,height=22cm}
\caption{}
\end{figure}

\newpage
\begin{figure}[h]
\psfig{file=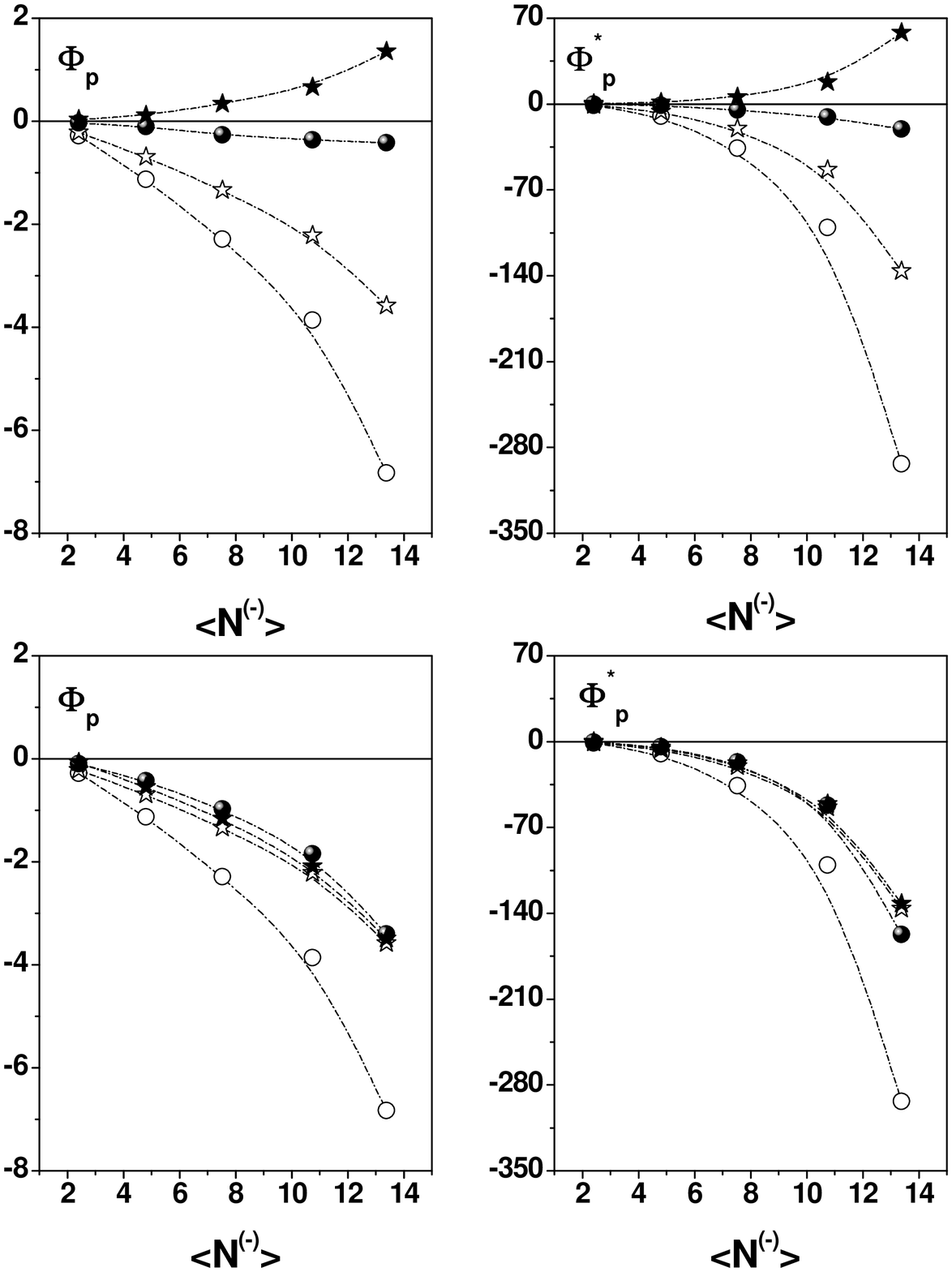,height=22cm}
\caption{}
\end{figure}
\newpage
\begin{figure}[h]
\psfig{file=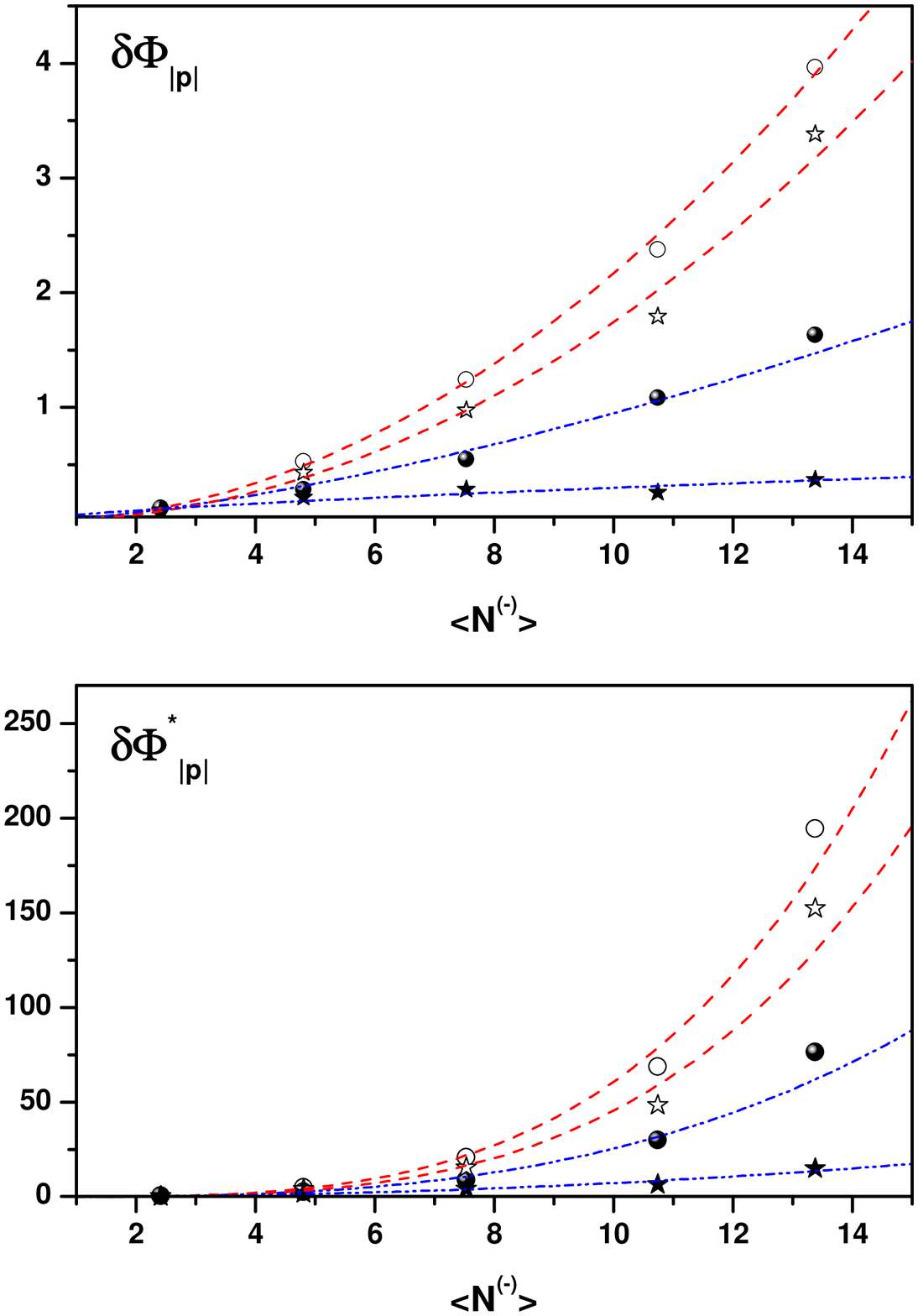,height=22cm}
\caption{}
\end{figure}

\begin{thebibliography}{99}
\bibitem{FOOT} If $AA$ collision is superposition of $NN$
                interactions with no secondary collisions, the value
                of $\Phi$ is exactly the same for the $NN$ and $AA$
                cases \cite{GM}.
\bibitem{FOOT2} Because so far our numerical algorithm for BEC is
                 available only for essentially one dimensional 
                 hadronization with $p_T$-dependence summarized
                 by transverse mass $\mu_T$, the $x=|p|$ variable is
                 the only choice we have for this kind of comparison.
\bibitem{FOOT3} In \cite{TIHANY} in addition to the constant weigths
                 arbitrary chosen as $P=0.5$, also a kind of the
                 "most natural" choice of weights was considered
                 (using only the information provided by event
                 generator) namely, for a pair of particles $(ij)$
                 weight $P(ij) = \exp\left[- \frac{1}{2}
                 \delta^2_{ij}(x)\cdot\delta^2_{ij}(p) \right]$
                 for CAS (particles $(ij)$ described by
                 spacio-temporal wave packets separated by
                 $\delta_{ij}(x)$ would have widths given by their
                 momentum separation $\delta_{ij}(p)$) and weight
                 $P(ij) =\exp\left[-\, \frac{\delta^2_{ij}(p)}{2\mu_T
                 T_l} \right]$ for MaxEnt ($\mu_T$ is transverse mass
                 put here to be equal $0.3$ GeV and the role of
                 spatial dimension is now played by the "temperature"
                 $T_l$ of the $l^{th}$ event).
\end{thebibliography}
\end{document}